\documentclass[a4paper,11pt]{article}
\usepackage{jinstpub} 
\usepackage{lineno}

\title{\boldmath Compact sub-10~ps Resolution Radio Frequency Photomultiplier Tube}

\author[a]{S.~Abrahamyan,}
\author[a,1]{S.~Zhamkochyan,\note{Corresponding author.}}
\author[a]{H.~Rostomyan}
\author[a]{A.~Margaryan,}
\author[a]{H.~Elbakyan,}
\author[a]{A.~Kakoyan,}
\author[a]{A.~Papyan,}
\author[a]{A.~Safaryan,}
\author[b]{J.~Annand,}
\author[b]{K.~Livingston,}
\author[b]{R.~Montgomery,}
\author[c]{S.N.~Nakamura,}
\author[d]{P.~Achenbach,}
\author[d]{J.~Pochodzalla,}
\author[e]{D.L.~Balabanski,}
\author[f]{V.~Sharyy,}
\author[f]{D.~Yvon,}
\author[g] {A.~Aprahamian,}
\author[a]{V.~Kakoyan}
\affiliation[a]{A.I.~Alikhanyan National Science Laboratory (Yerevan Physics Institute),\\
2 Alikhanyan Brothers., Yerevan, Armenia}
\affiliation[b]{School of Physics \& Astronomy, University of Glasgow,\\
G12 8QQ Scotland, UK}
\affiliation[c]{Department of Physics, Graduate School of Science, the University of Tokyo,\\
Tokyo, Japan}
\affiliation[d]{Institut für Kernphysik, Johannes Gutenberg-Universität Mainz,\\
Mainz, Germany}
\affiliation[e]{Extreme Light Infrastructure- Nuclear Physics (ELI-NP),\\
Bucharest-Magurele, Romania}
\affiliation[f]{Département de Physique des Particules Centre de Saclay,\\
91191 Gif-sur-Yvette Cedex, France}
\affiliation[g]{Department of Physics and Astronomy, University of Notre Dame,\\
Notre Dame, IN 46556, USA}

\emailAdd{szh@mail.yerphi.am}

\abstract{
Experimental measurements of the radial spreading of photoelectrons emitted from a multi-alkali photocathode in a MCP-based photomultiplier tube have shown that, for photon wavelengths of 455~nm, 515~nm and 625~nm, the maximum initial energies of the emitted photoelectrons are approximately 0.3~eV, 0.2~eV and 0.1~eV respectively. Combining these experimental results with simulations performed using the SIMION simulation package, a compact radio-frequency photoelectron multiplier tube with a temporal resolution better than 10 ps is proposed. The device would be suitable for applications in several fields, particularly in medical optical instruments employing time-correlated single-photon counting.
}


\begin{document}
\maketitle
\flushbottom

\section{Introduction}

The development of photon detectors with ever-improving temporal resolution remains a central challenge in modern experimental science and applied photonics. Applications such as time-correlated single-photon counting (TCSPC), time-of-flight measurements, positron emission tomography, ultrafast spectroscopy, quantum optics, and optical medical diagnostics require detectors that combine high sensitivity, low dead time, and picosecond level timing precision. Although substantial progress has been achieved with microchannel plate photomultiplier tubes (MCP-PMTs), silicon photomultipliers (SiPMs), and hybrid photodetectors, achieving reliable sub-10 ps timing resolution -- particularly at long wavelengths in the red and near infrared spectral regions -- remains a significant challenge (see~\cite{spd_Eisaman_2011,spd_Hadfield_2009,spd_EsmaeilZadeh_2017} and references therein). In recent years, superconducting nanowire single-photon detectors (SNSPDs) have demonstrated single-photon timing resolution below 10~ps~\cite{snpd_Korzh_2020}. However, their operation requires liquid-helium temperatures, which introduces additional technical complexity.

In conventional vacuum photomultiplier based systems, the ultimate timing resolution is limited by several factors, including the intrinsic photoelectron emission process at the photocathode, the transit-time spread during electron multiplication, and bandwidth of the readout electronics. Radio-frequency (RF) time-to-space conversion techniques provide an alternative approach to high-precision timing. In such systems, the arrival time of electrons is encoded into a spatial coordinate using a high-frequency electromagnetic field. Streak cameras based on RF deflection routinely achieve picosecond and sub-picosecond temporal resolution. However, their application to single-photon counting is limited due to complex optical readout and restricted count-rate capability. 

The Radio-Frequency Photo Multiplier Tube (RFPMT) concept combines the sensitivity of photocathode based photon detection with RF-based ultra fast time encoding. In this approach, photoelectrons are accelerated to keV energies, focused by means of an electrostatic lens and passed through an RF deflector, which converts their temporal distribution into a position distribution on a circular scan. A position-sensitive detector, typically based on microchannel plates coupled to a position-sensitive anode, provides fast electronic readout with minimal dead time. This architecture enables picosecond level precision while remaining compatible with single-photon counting at high repetition rates~\cite{rftimer_jinst,rftimer_nim}.

In this work, we investigate the fundamental limitations imposed by the photoelectron emission process in a multialkali photocathode operating at different photon wavelengths. Detailed electron-optical simulations were performed with SIMION~\cite{simion} using photo-emission parameters derived from the experimental results. The simulations demonstrate the feasibility of a compact RFPMT configuration, without any electrostatic focusing, capable of achieving temporal resolution better than 15~ps, particularly for long wavelength photons in the red and near-infrared regions, where resolution below 10~ps is achievable. The proposed detector architecture is well suited for Time Correleated Single Photon Counting (TCSPC)-based optical medical devices, where compactness, high timing precision, and low dead time are essential.

\section{Measurements}
\label{sec:measurements}

\begin{figure}[th]
\centering
\includegraphics[width=0.8\textwidth]{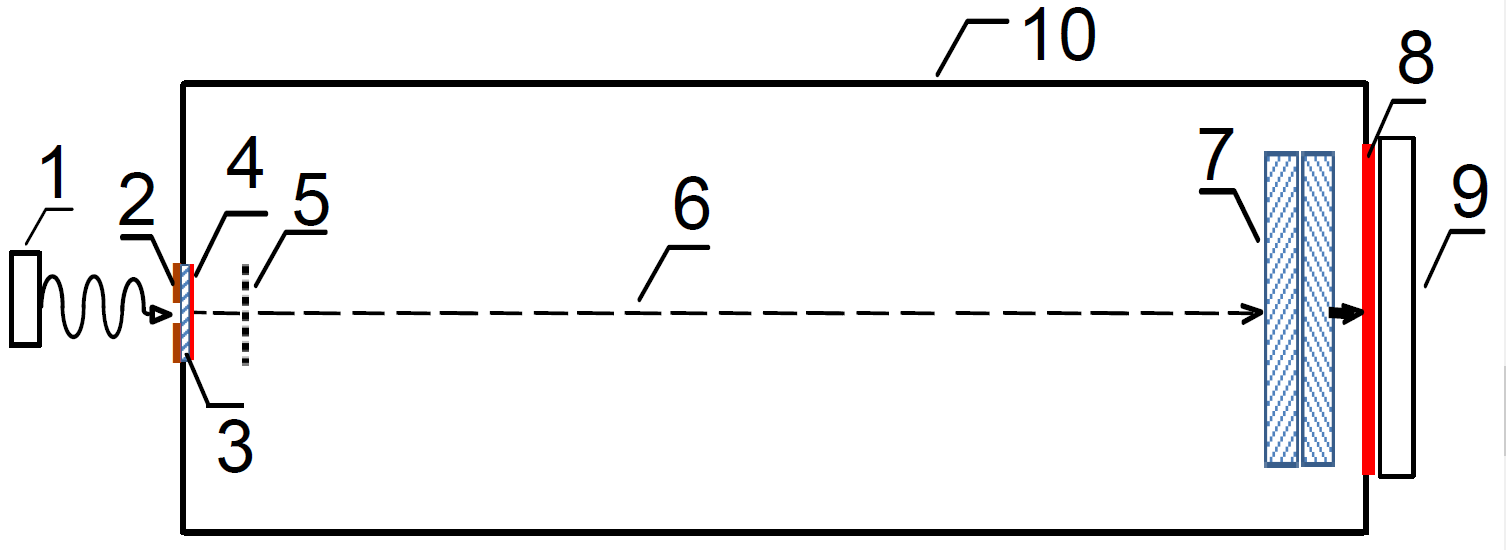}
\caption{Schematic of the experimental setup. 1~-~photon source; 2~-~collimator; 3~-~quartz window; 4~-~photocathode; 5~-~accelerating mesh; 6~-~photoelectron; 7~-~MCP detector; 8~-~resistive layer; 9~-~delay-line anode, 10~-~vacuum tube.}
\label{fig:exp_setup}
\end{figure}

Experimental investigations were performed using the setup schematically illustrated in Fig.~\ref{fig:exp_setup}. LED light sources with wavelengths of 455~nm, 515~nm, 625~nm were used to illuminate a photocathode manufactured by Photek~\cite{cathode_photek} through a collimator aperture of 0.9~mm diameter. The photoelectrons emitted from the photocathode were accelerated over the distance of 2 mm and subsequently propagated through a 31~cm drift region before reaching the face of the position-sensitive detector (PSD). The latter consists of an MCP-based electron multiplier, a resistive layer at the rear of the vacuum tube, and a charge sensitive delay line anode mounted externally (air-side). The spatial resolution of the PSD was approximately $\sigma\sim300\mu m$. The data acquisition system was based on PicoScope \cite{picoscope}.

\begin{figure}[ht]
\centering
\includegraphics[width=0.95\textwidth]{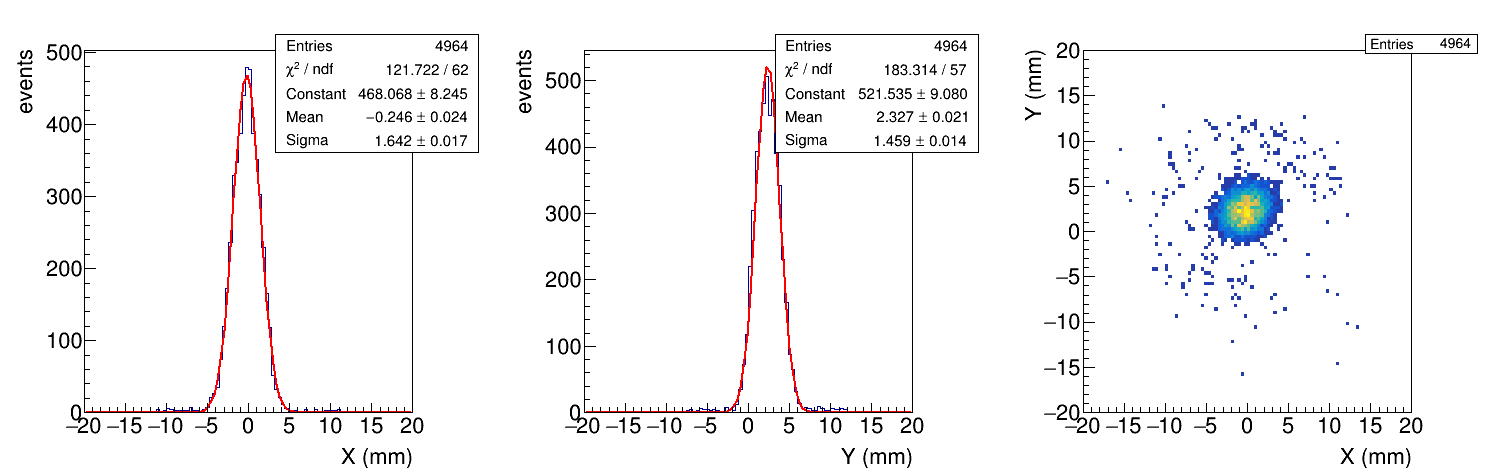}
\vspace{4mm}
\includegraphics[width=0.95\textwidth]{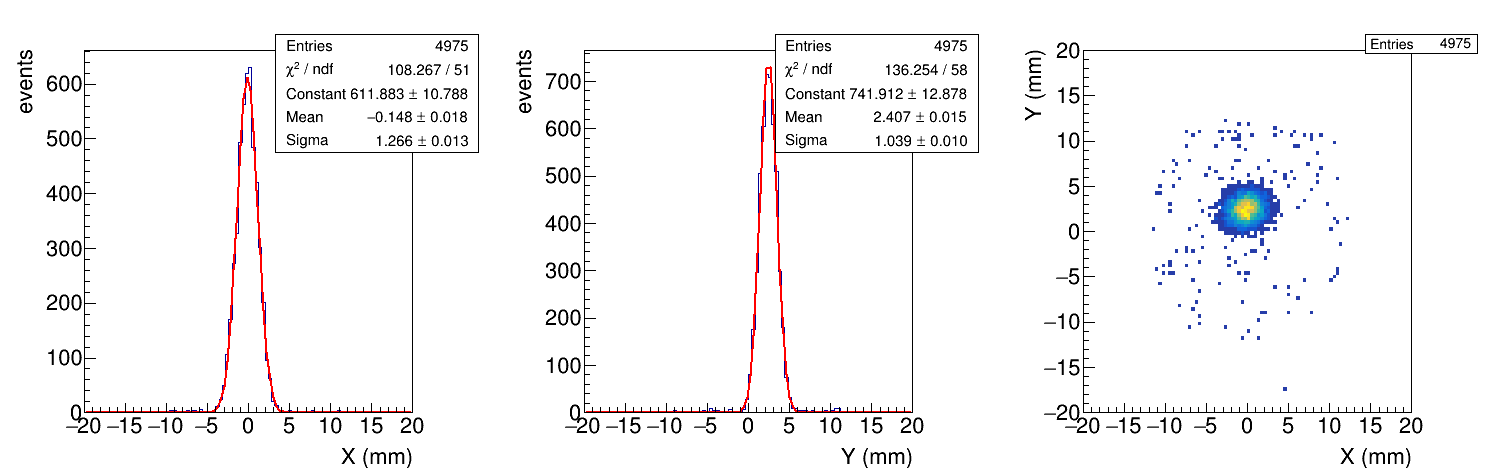}
\vspace{4mm}
\includegraphics[width=0.95\textwidth]{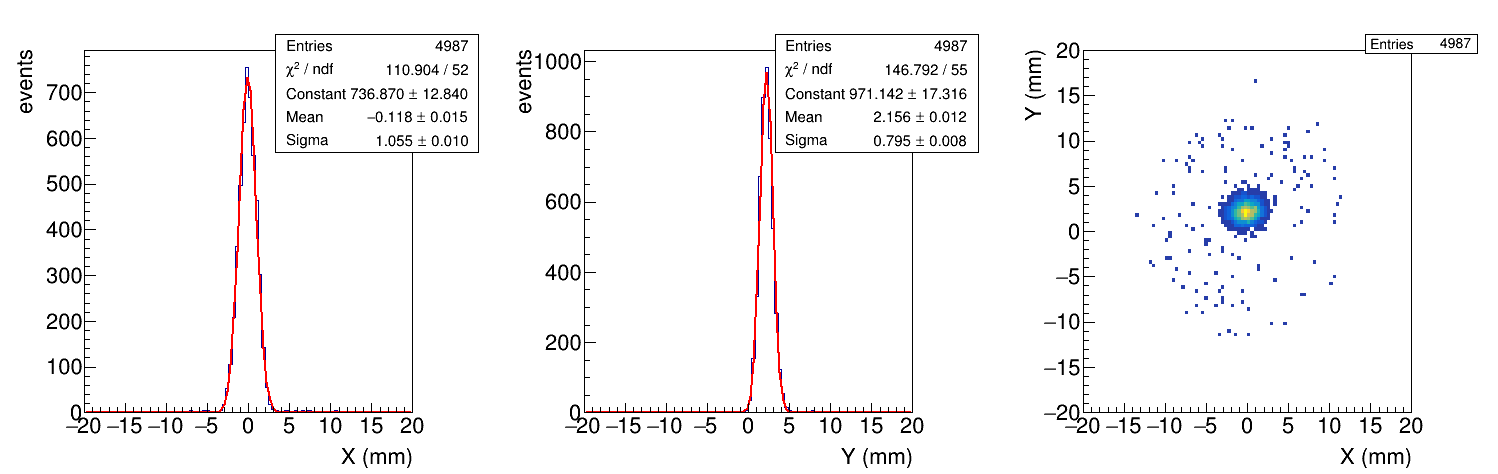}
\caption{Measured secondary-electron transverse beam coordinates distributions for 455~nm~(top), 515~nm~(middle) and 625~nm~(bottom) illumination. For each wavelength the X projection (left), Y projection (center), and two-dimensional Y vs X distribution (right) are shown.}
\label{fig:blue_red}
\end{figure}

Figure~\ref{fig:blue_red} shows representative distributions of transverse beam coordinates, revealing a pronounced wavelength dependence. To estimate the photoelectron initial energy distribution corresponding to each wavelength, simulations using the SIMION software were performed. The initial electrons were created uniformly distributed over a 0.9~mm diameter area. An isotropic angular distribution and an initial kinetic energy uniformly distributed between 0 and a value $E_{max}$ were assumed. Electrons were propagated through exactly the same geometry and electric fields as in the experimental setup and their positions at the detector face were obtained. By varying the value of $E_{max}$ until the resulting position distribution of the simulated electrons matched the one obtained experimentally (Fig.~\ref{fig:blue_red}) the energy distributions for each studied wavelength were estimated.

The best-match values of the maximum initial energy of photoelectrons are 0.3~eV at 455~nm; 0.2~eV at 515~nm and 0.1~eV at 625~nm, in agreement with previously reported results for multi-alkali (Sb–K–Na–Cs) photocathodes~\cite{photoelectrons_Eberhardt_1977}. Based on these findings we propose the development of a compact RFPMT with a simplified electron-optical structure.


\section{Compact RFPMT design concept}
\label{sec:rfpmt}

\begin{figure}[ht]
\centering
\includegraphics[width=0.6\textwidth]{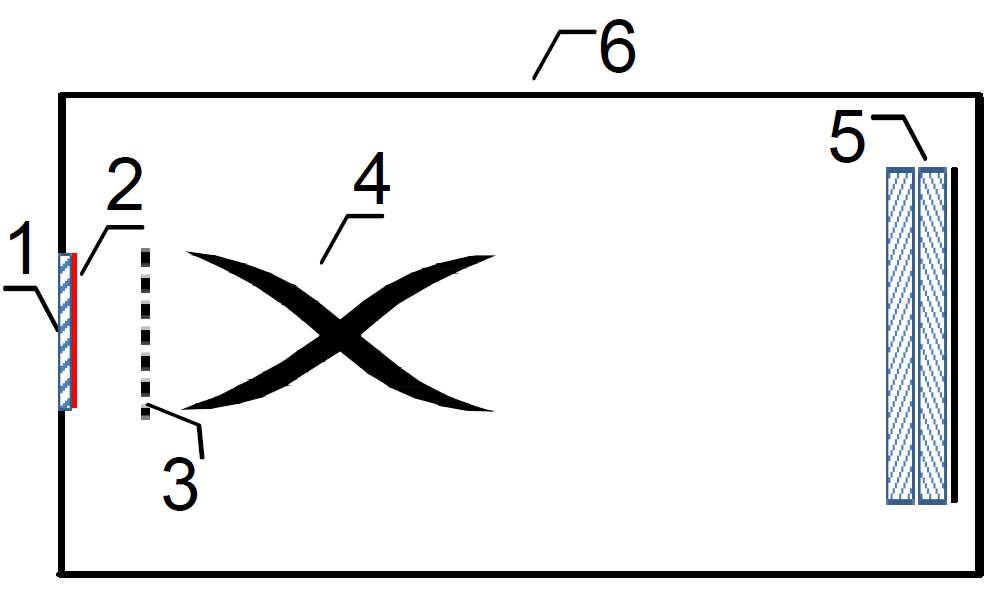}\caption{Schematic of the compact RFPMT: 1~-~quartz window, 2~-~photocathode, 3~-~accelerating mesh, 4~-~RF deflector, 5~-~position-sensitive detector, 6~-~vacuum tube.}
\label{fig:rfpmt_schema}
\end{figure}

A schematic of the proposed compact RFPMT is shown in Fig.\ref{fig:rfpmt_schema}. Photons enter the tube through a quartz window, and photoelectrons are emitted from a photocathode biased at -2500 V. The electrons are accelerated across a short (2~mm) gap to a grounded electrode and subsequently transported through an RF deflector. RF-scanned electrons are detected by a MCP-based position-sensitive detector located 12 cm from the photocathode. The azimuthal angle of the deflected photoelectron on the detector's face depends on the photoelectrons time with respect to the phase of the RF power applied to the deflector. Hence, by measuring the position and determining the angle - one obtains the photoelectron time~\cite{rftimer_theory}. 

Three terms  contribute to the timing resolution.
\begin{itemize}
  \item The time dispersion of photoelectrons while exiting the photocathode. This is typically below 1~ps and is neglected in the following discussion.
  \item The photoelectron transit time spread ($\sigma_{TTS}$), defined as the variation in the transit time of photoelectrons from their emission at the photocathode to their exit from the deflector.
  \item The technical time resolution which depends on the transverse beam size and spatial resolution of the position-sensitive detector, frequency of the RF source, and opening radius of the deflection.
\end{itemize}
Hence, the resulting time resolution is given by the following equation.
\begin{equation}
\sigma_t = \sqrt{\left(\frac{\sigma_r}{2\pi R f}\right)^2+\sigma_{TTS}^2},
\label{eq:time_resolution}
\end{equation}
where $\sigma_t$ is the resulting timing resolution in ns, $\sigma_r$ is the measured transverse size of the beam at the detector plane, $R$ is the RF scan radius, $f$ is the RF frequency in Hz and $\sigma_{TTS}$ is the sigma of the photoelectrons transit time distribution. $\sigma_r$ is in turn determined by the beam spread ($\sigma_{beam}$) and the resolution of the position-sensitive detector ($\sigma_{psd}$).

\begin{equation}
\sigma_r = \sqrt{\sigma_{beam}^2+\sigma_{psd}^2}
\label{eq:spot_size}
\end{equation}

To estimate achievable timing performance in a compact RFPMT, SIMION based Monte Carlo (MC) simulations of the setup presented in Fig.~\ref{fig:rfpmt_schema} were performed. Electrons uniformly distributed on a photocathode exit with an isotropic angular distribution and uniformly distributed energy from 0 to $E_{max}$. The electrons are accelerated in the 2~mm region region between cathode and mesh to 2.5 keV and then propagated to the screen located at 12~cm, where their positions and transit times are recorded. MC simulations were performed for values of $E_{max}$: 0.3~eV, 0.2~eV and 0.1~eV.  which, according to the results presented in Section~\ref{sec:measurements}, correspond to 455~nm, 515~nm and 620~nm, respectively.

\begin{figure}[th]
\centering
\includegraphics[width=0.75\textwidth]{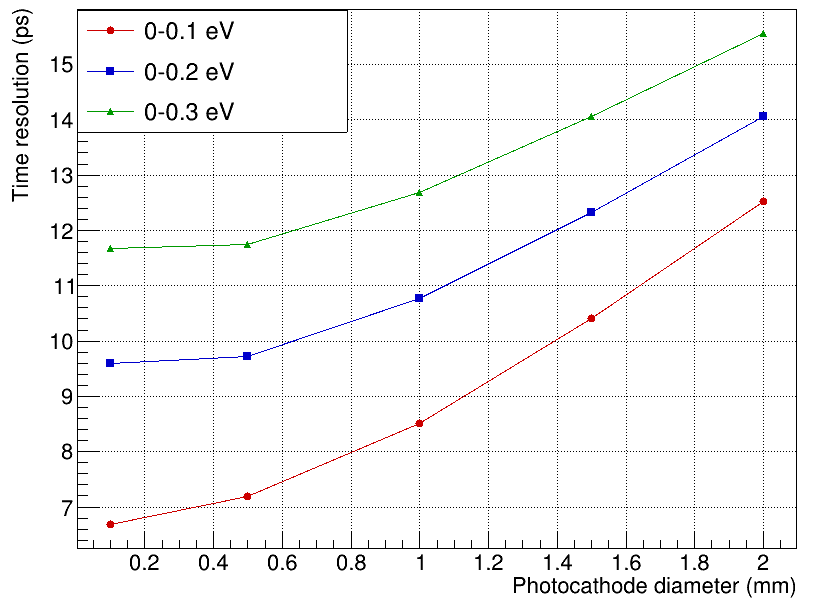}
\caption{Estimated time resolution vs photocathode size for three electron energy spread scenarios: $\Delta E = 0-0.1$~eV (red circles), $\Delta E = 0-0.2$~eV (blue squares), and $\Delta E = 0-0.3$~eV (green triangles).}
\label{fig:sim_results}
\end{figure}

The resulting position and time distributions are then used to estimate the time resolution based on equation~\ref{eq:time_resolution}. The photocathode size dependence of the resulting time resolution for three values of $E_{max}$, a scan radius of $R=15$~mm and RF frequency $f=500$~MHz are presented in Fig.~\ref{fig:sim_results}. The results indicate that a time resolution $\lesssim$15~ps is expected for photocathode diameters up to 2.0~mm, while a resolution of 10~ps or better is achievable for longer wavelengths and smaller photocathode sizes.

\section{Conclusions}

In this work, we have experimentally investigated the photoelectron emission characteristics of a multialkali photocathode over a wavelength 455-625~nm and quantified the wavelength dependency of the initial energy spread of the emitted photoelectrons. By incorporating the experimentally determined emission parameters into SIMION simulations, we have shown that a compact radio frequency photomultiplier tube, without an electrostatic focusing lens, can achieve temporal resolution better than 15 ps and approach the sub-10 ps regime for longer wavelength photons. The proposed RFPMT architecture combines keV-scale photoelectron acceleration, RF time-to-space conversion, and fast position sensitive readout, enabling high timing precision with low dead time in a simplified and compact detector geometry. The expected performance and compact design make the proposed RFPMT particularly attractive for time-correlated single photon counting applications in optical medical devices, including fluorescence lifetime imaging and time-resolved diffuse optical techniques. 

\acknowledgments
This work was supported by the Higher Education and Science Committee of the Republic of Armenia (Research Project: 23LCG-1C018) and the International Science and Technology Center (ISTC project AM-2803).

\bibliographystyle{JHEP}
\bibliography{biblio}

\end{document}